# An Explanation From First-Principle Equations For The Universality of Non-Gaussian Distributions in Edge Plasma Fluctuations

F. Sattin

*Consorzio RFX, Associazione EURATOM-ENEA sulla fusione,*
*Corso Stati Uniti 4, Padova, ITALY*

**Abstract**

Probability Distributions Functions (PDFs) of fluctuations of plasma edge parameters are skewed curves fairly different from normal distributions, whose shape appears almost independent of the plasma conditions and devices. We start from a minimal fluid model of edge turbulence and reformulate it in terms of uncoupled Langevin equations, admitting analytical solution for the PDFs of all the fields involved. We show that the supposed peculiarities of PDFs, and their universal character, are related to the generic properties of Langevin equations involving quadratic nonlinearities.

       52.65.Kj     MHD and fluid simulation

The fluctuations of practically any quantity measured at the edge of plasma devices are intermittent. Qualitatively, this means that quite frequent strong bursts—far exceeding the average amplitude--are observed. The high-amplitude bursts are commonly interpreted as due to the collective motion of finite volumes of plasma, and represent the nonlinear saturated state of some instability. The resulting transport is accordingly affected by their presence: it departs from a purely diffusive picture, valid in the Gaussian case of small independent fluctuations, and acquires instead a convective component associated to the rigid displacement of these coherent structures. Since edge transport critically affects the performances of the magnetic fusion devices, it is fundamental to fully understand its physics [1,2].

Experimentally, most of the investigations are carried out through single-point or few-points measurements, which produce results in terms of scalar timeseries. The existence of intermittence is diagnosed from the statistical properties of the signals. The Probability Distribution Function (PDF) quantifies the relative frequency of appearance of the signal at a given amplitude, and is one among the most used statisical measures. Experimental PDFs do commonly deviate from Gaussian curves. This is strikingly clear for plasma density fluctuations, which feature a highly skewed shape with an almost exponential tail, see, e.g., [2,3]. (Most investigations are carried on using Langmuir probes that actually measure a saturation current $J_{sat}$. Since $J_{sat} \propto n \times T^{1/2}$ and temperature fluctuations are usually smaller than density ones, the distinction between $J_{sat}$ and $n$ is not always stated). Virtually all other signals, e.g. electrostatic potential or temperature, are not Gaussian, but the degree of departure from normality is usually more limited (see, e.g., [4]). The fact that density PDFs are strongly skewed curves has thus attracted considerable attention, together with the evidence that qualitatively very similar PDFs—to a good extent—are encountered in all devices, irrespective of geometry or magnetic topology [2,3]. It is just natural to speculate that only some non-generic mechanism of transport can produce such peculiar PDFs, and furthermore it must exist in all plasmas because of the observed universal behaviour of PDFs.

Numerical codes have already shown to be able to succesfully tackle several aspects of the edge turbulence [5-7]. In this work we propose a simplified analytical treatment: most features from a full modelization are retained while at the same time an explicit solution for the PDFs is provided. The analysis of the structure of our analytical theory reveals that the exact spatial-temporal structure of the turbulence does not play any role in determining the PDFs. Their Gaussian character (or lack of) arises simply from the presence of nonlinear coupling between the turbulent fields, which is dictated by the physics of the problem and in our case is the simplest conceivable, namely a quadratic coupling. We finally validate the analytical theory by solving numerically the full set of equations, and show how the two solutions perfectly overlap.

Our starting point is a minimal fluid model for electrostatic interchange turbulence that involves evolution equations for the two fields density $n$ and vorticity $\Omega$. We consider a slab 2-dimensional geometry, with the $x$ axis playing the role of radial coordinate **r**, the $y$ axis the **r**×**B** direction, and the $z$ axis the direction parallel to the magnetic field **B**. The continuity equations for particle and charge density write respectively

$$\frac{dn}{dt} + nC(\phi) - C(nT) = -\frac{n}{\tau_n} \qquad (1)$$

$$\frac{d\Omega}{dt} - C(nT) = -\frac{\Omega}{\tau_\Omega} \qquad (2)$$

The symbol $d/dt$ stands for the advective derivative including the **E**×**B** drift: $d/dt = \partial/\partial t + B^{-1}\hat{\mathbf{z}} \times \nabla\phi \cdot \nabla$, with $\phi$ electrostatic potential; $C$ is a differential operator accounting for compressibility. It arises because of spatial variations of the magnetic field induced by curvature. In our slab geometry it writes $C = -\zeta \partial/\partial y$, with $\zeta$ a parameter of order of (Larmor radius/Major radius) quantifying the amount of curvature. Vorticity and potential are related through $\Omega = \nabla_\perp^2 \phi$, where the pedix means differentiation with respect to $x$ and $y$. The terms on the r.h.s. are frictions, e.g. driven by neutrals [8]. Temperature $T$ is constant: $T = 1$ in dimensionless units. A version of this model has been extensively studied, e.g., in [6,7], where it was shown that it successfully works in reproducing statistical features of real devices.

This set of deterministic coupled partial differential equations must be simplified in order to be analytically tractable. The first and main simplification consists in artificially decoupling the equations. For that, we label fields into each equation into "main" and "auxiliary" fields. Main field is the field that is explicitly advanced in time in each equation; auxiliary fields are all the other fields occurring. Thus, in Eq. (1) the main field is the density $n$, and the auxiliary field the potential $\phi$. In Eq. (2), the main field is $\Omega$ and the auxiliary fields $n$ and $\phi$. The next step is to replace into each equation the exact self-consistent dynamics of the auxiliary fields with a prescribed one, conveniently chosen. We are dealing with turbulent quantities, and a close surrogate for a turbulent dynamics is a stochastic one. Hence, we replace auxiliary fields with stochastic fields (white noises). This turns Eqns (1,2) into decoupled stochastic differential equations. The analytical solution of the associated Fokker-Planck Equations yields the sought PDFs.

Let us start from Eq. (1). It is actually an equation for $\psi = \ln(n/\tilde{n}_0)$, with $\tilde{n}_0$ a normalization constant. After some rearrangements, we may thus write

$$\frac{\partial \psi}{\partial t} = -\frac{\partial \psi}{\partial y}\frac{\partial \phi}{\partial x} + \frac{\partial \psi}{\partial x}\frac{\partial \phi}{\partial y} - \zeta \frac{\partial \phi}{\partial y} - \zeta \frac{\partial \psi}{\partial y} - \frac{1}{\tau_n} = -\{\phi,\psi\} - \zeta \hat{y}\cdot\nabla\phi - \zeta \hat{y}\cdot\nabla\psi - \frac{1}{\tau_n} \qquad (3)$$

If we replace $\phi$ with a white-noise stochastic variable, we choose to ignore the exact spatial-temporal correlations that $\phi$ must fulfil because of the dynamics (2)--and that enter into (1) because of the presence of spatial derivatives $\partial_{x(y)}\phi$. This has obvious consequences wherever the exact correlations are needed (for example, in the computation of power spectrum); our guess is that discarding this information will be not detrimental to our goal. Of course, we are at the same time forgetting feedback effects, since $\phi$ is now considered a field given, unaffected by $\psi$. Since we are replacing in (1) the precise spatial correlations with approximate ones as long as the terms in $\phi$ are considered, there is no point in retaining them in full form for the field $\psi$; hence, we will replace its spatial derivatives with typical values of the field $\psi$ divided by some characteristic length:

$$\{\phi,\psi\} \approx \frac{\phi}{\lambda^2}\psi \equiv -\tilde{\eta}\psi; \quad -\zeta \hat{y}\cdot\nabla\phi \equiv \tilde{\xi}; \quad -\zeta \hat{y}\cdot\nabla\psi \approx -\frac{\zeta}{\lambda}\psi \equiv -\tilde{\nu}\psi \qquad (4)$$

Hence, Eq. (3) becomes

$$\frac{d\psi}{dt} = \tilde{\eta}\psi + \tilde{\xi} - \tilde{\nu}\psi - \frac{1}{\tau_n} \qquad (5)$$

which ultimately writes

$$\frac{d\psi}{dt} = \eta\psi + \xi - \nu\psi \qquad (6)$$

after the positions $\psi \to \psi - \psi_0, \tau_n^{-1} = -\tilde{\nu}\psi_0, \psi_0 = \ln(\tilde{n}_0/n_0), \nu = \tilde{\nu}, \eta = \tilde{\eta}, \xi = \tilde{\xi} + \eta\psi_0$.

Since they are defined in terms of $\phi$, $\xi$ and $\eta$ are stochastic functions as well. Eq. (6) is thus a Langevin equation containing both additive ($\xi$) and multiplicative ($\eta$) noise, together with a friction term ($\nu$), which must be positive in order for the solution to be bounded. It ultimately may be taken equal to unity, since amounts to rescaling time. Furthermore, $\xi$ and $\eta$ cannot be uncorrelated, since both functions of the potential $\phi$; hence the correlations functions for the noises $\xi$ and $\eta$ are

$$\begin{aligned}\langle \eta(t)\eta(t')\rangle &= \alpha^2 \delta(t-t') \\ \langle \xi(t)\xi(t')\rangle &= w^2 \delta(t-t') \\ \langle \eta(t)\xi(t')\rangle &= \alpha w \varepsilon \delta(t-t'), \quad \varepsilon \neq 0\end{aligned} \qquad (7)$$

Using standard methods [9] we may derive from (6,7) a Fokker-Planck Equation (FPE) for the probability $p_\psi(\psi)$ (A technical aside: drift and diffusion coefficients of the FPE may be computed according to two conventions: either Itô or Stratonovich [9]. We follow here Itô's but our conclusions ultimately do not depend upon the precise choice done). At steady state, FPE writes

$$0 = \frac{d}{d\psi}\{v\psi\, p_\psi(\psi)\} + \frac{1}{2}\frac{d^2}{d\psi^2}\left[\left(\alpha^2\psi^2 + 2\varepsilon\alpha w\psi + w^2\right)p_\psi(\psi)\right] \quad (8)$$

whose solution is

$$p_\psi(\psi) \propto \frac{\exp\left\{\dfrac{2\varepsilon}{\alpha^2\sqrt{1-\varepsilon^2}}\arctan\left(\dfrac{\alpha\psi+\varepsilon w}{w\sqrt{1-\varepsilon^2}}\right)\right\}}{\left(\alpha^2\psi^2 + 2\varepsilon\alpha w\psi + w^2\right)^{(1+\alpha^{-2})}} \quad (9)$$

which may eventually be written as a PDF for the density, $p_n(n)$, through a simple change of variables:

$$p_n(n) = N \times p_\psi\left(\psi = \ln\left(\frac{n}{n_0}\right)\right) \times \frac{n_0}{n} \quad (10)$$

(Where $N$ is a normalization factor).

Let us now repeat the same steps in Eq. (2), where now we invert the roles and consider $n$ and $\phi$ as stochastically fluctuating variables and $\Omega$ the variable we wish to solve for. It is straightforward to reach a Langevin equation formally identical to (6), but with a new set of coefficients ($\eta'$, $\xi'$, $\nu'$):

$$\frac{d\Omega}{dt} = \eta'\Omega + \xi' - \nu'\Omega \quad (11)$$

The stationary PDF from (11) is again given by an expression of the form (9) where now the role of $\psi$ is played by $\Omega/\Omega_0$. However, at the quantitative level we may expect some differences to arise between the two solutions: in (6) $\eta$, $\xi$ are both functions of the derivatives of the potential. Hence, we expect the two stochastic variables to be highly correlated ($|\varepsilon| \sim O(1)$). On the contrary, $\eta'$ and $\xi'$ involve two different quantities: respectively, the potential and the density. We expect therefore a lesser degree of correlation. If we take the limiting case $\varepsilon = 0$ in (9), the resulting PDF is $p(x) = \left[\alpha^2 x^2 + w^2\right]^{-\left(1+\frac{1}{\alpha^2}\right)}$. That is, the PDF has zero skewness and fat algebraic tails.

Before going further, we comment on the passages leading from Eq. (3) to (6). They may appear completely arbitrary but, eventually, some formal basis may be provided for them. First of all, we point out that an experimental time series should not be compared against one single numerical simulation, rather is closer to an average over several independent runs. The reason is that, experimentally, there will unavoidably be unmodelled perturbations acting as decorrelation mechanisms. The simulation should be therefore stopped after each decorrelation time and restarted with a set of new initial conditions. The averaging done in Eqns. (7,8) amounts to replacing the deterministic fields arising from one numerical realization with an average done over several uncorrelated ones, with the stochastic element ultimately coming from the arbitrariness of the perturbation due to the background. The caveat that numerical and experimental quantities should be compared only after the former ones have been adequately averaged over a meaningful set of realizations, has been addressed in recent publications [10]. We point out also the recent paper by Materassi and Consolini [11] that shares many affinities with the spirit of the present work.

Our guesses above look admittedly very rough. It is necessary to confirm them against full numerical simulations. Therefore, we studied numerically the set (1-2). The geometry chosen was like in Figs. (1,2): a rectangular box of size $L_x = 64$, $L_y = 128$. Periodic boundaries conditions were set at $y = 0$ and $y = L_y$. The edge at $x = L_x$ was treated as an absorbing boundary: $n = \Omega = \phi = 0$. At the other boundary we imposed all the fields to be linear combinations of travelling waves:

$$F = F_0 + \sum_k A_k \cos(kx - w_k t + \varphi_k) \qquad (12)$$

where $F = n, \Omega, \phi$. In these simulations we used three waves with $k_1 = 4\pi L_y^{-1}$, $k_2 = (3/2)k_1$, $k_3 = (7/2)k_1$ and $w_k = 2k$. The values of the other coefficients are

$$\begin{cases} n: F_0 = 1;\ A_{1,2,3} = 1;\ \varphi_{1,2,3} = \pi/2 \\ \Omega: F_0 = 0;\ A_{1,2,3} = -k_{1,2,3}^2;\ \varphi_{1,2,3} = 0 \\ \phi: F_0 = 0;\ A_{1,2,3} = 1;\ \varphi_{1,2,3} = 0 \end{cases} \qquad (13)$$

The numerical value of the parameter appearing in the curvature term is $\zeta = 0.1$. This value may appear anomalously large, but note that, by virtue of the structure of Eqns. (1,2), it is possible to vary $\zeta$ provided we correspondingly rescale the times. Finally, a small diffusivity ($D = 10^{-2}$) for both $n$ and $\Omega$ is included for numerical stability purposes. The choice (12,13) is not critical to our purposes, but is a convenient way to provide a seed to growing perturbations. It is inspired by the suggestion from Mattor and Diamond that the radial propagation of drift waves might be at the basis of edge turbulence [12].

As solver we used the powerful finite-elements commercial program COMSOL [13], which allows for a very fast and efficient programming. The mesh used is displayed in Fig. (1). We made the system to start with all fields set to very low levels and left it to evolve up to large times ($t_{max} = 1.73 \times 10^5$). A snapshot of the $n$ profile is shown in Fig. (2). After a few hundreds time units all fields converge towards statistical equilibrium, and from there onwards data were recorded at intervals of two units of time at a set of ($x,y$) locations, and binned in order to build the corresponding PDFs.

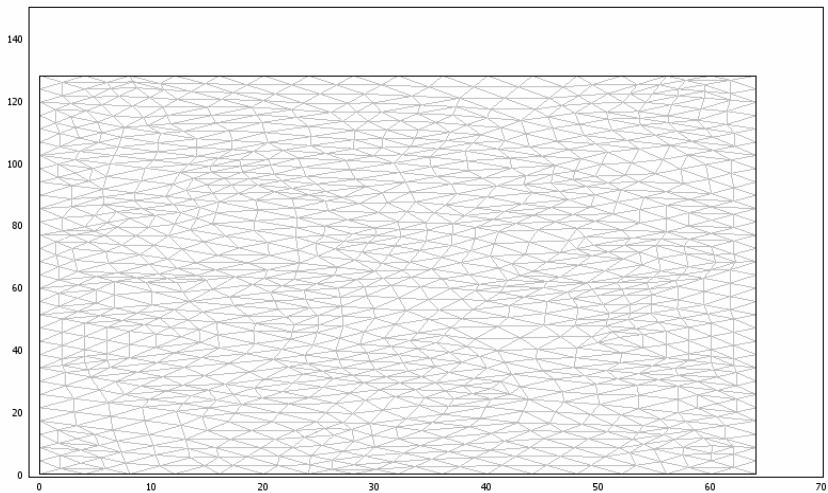

**Figure 1**. Mesh used in the computations.

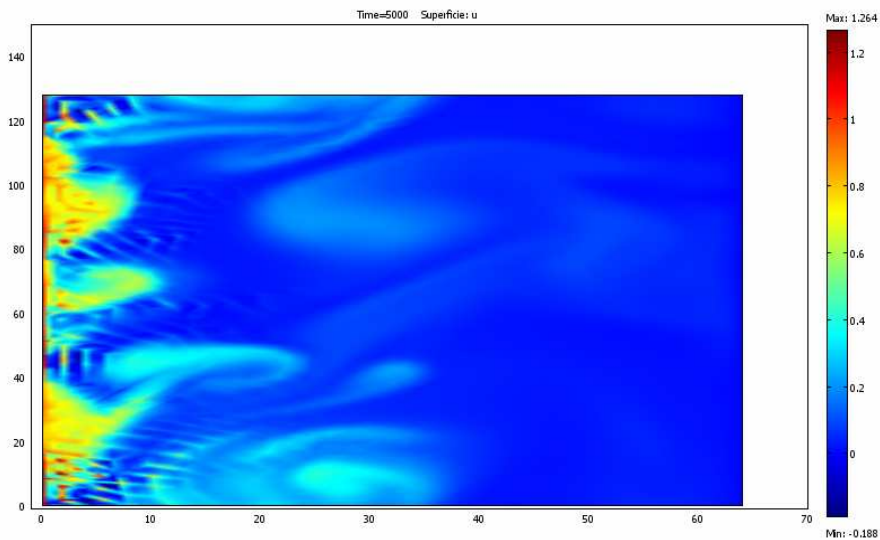

**Figure 2**. Snapshot of the density profile at $t = 5000$.

In Fig. (3) we present the first fundamental result of this paper. Dots are the PDF of density values at one spatial locations as computed by the numerical code. They are interpolated by curve (10). The agreement looks quite good. In Fig. (4) we report the same for the vorticity field. Again, the agreement is good, although numerical data show some scatter. It is remarkable that, like anticipated earlier, vorticity PDF is fitted with the curve (9) by using a smaller cross-correlation coefficient ε.

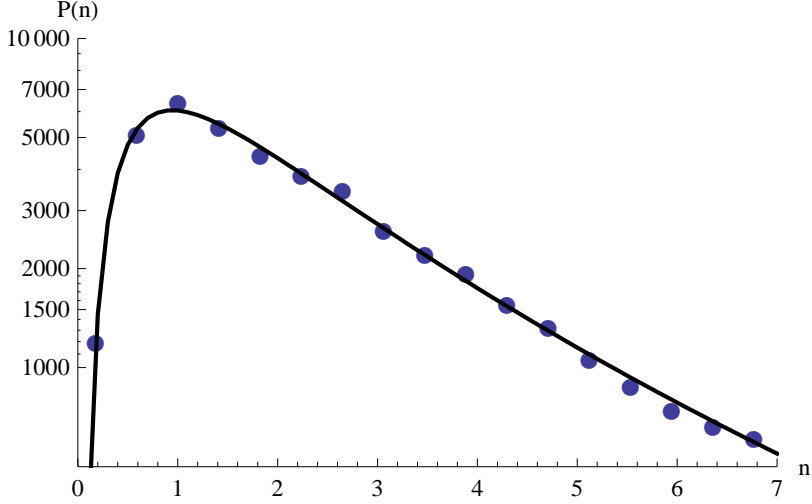

**Figure 3**. PDF of density fluctuations taken at the point $(x,y) = (40,60)$. Symbols, numerical results; solid line, best fit done using the analytical curve (10). Density has been rescaled so that the maximum of the PDF lies at $n = 1$. Best fitting parameters are: $n_0 = 2.17$, $a = 0.007$, $w = 1.29$, $\varepsilon = -0.47$.

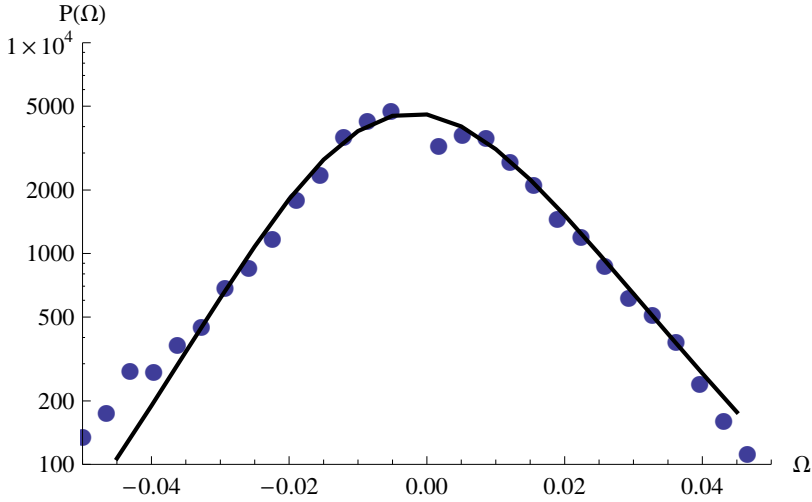

**Figure 4**. PDF of vorticity fluctuations taken at the point $(x,y) = (40,60)$. Symbols, numerical results; solid line, best fit done using the analytical curve (9). Best fitting parameters are: $\Omega_0 = -0.069$, $a = 0.65$, $w = 0.317$, $\varepsilon = -0.207$.

The results above provide also an insight about some conclusions from earlier studies. Let us consider the case when, due to some constraint or by choice, we have to linearize Eq. (3) around some stationary (non fluctuating) field. This amounts to discarding the $\eta\psi$ term in (6). It is then easy to check that the corresponding Fokker-Planck solution is a Gaussian, and one recovers the adiabatic-electrons-case discussed earlier. The papers [6] present an extensive study of the statistics of fluctuations in a slab geometry like ours. One of their results is that the density PDF acquires more and more a gaussian character in going from the edge toward the core boundary. Interestingly, in [6], core boundary conditions were set as $\partial\phi/\partial y = \partial\Omega/\partial y = \partial n/\partial y = \partial n/\partial x = 0$, i.e., exactly the conditions that annihilate the electric drift term. Therefore, our results suggest that the change of the shape of the PDF in [6] is not due to a modification to the nature of the turbulence but arises because of the boundary conditions imposed. A similar speculation may be put forth in connection with papers [14]. Those works carried on three-dimensional simulations using frozen profiles, which practically amounts to linearizing Eq. (3) but for retaining the nonlinear **E×B** term. Unlike studies carried on using the full nonlinear models, these papers found only Gaussian PDFs. We may speculate that the linearization of Eq. (3) forces fluctuations to be small and therefore the quadratic **E×B** term to be dominated by the others. It is further interesting to notice that Ref. [15] contains one of the first attempts of analytically interpolating experimental PDFs, in that occasion in terms of a log-normal curve. This result may be straightforwardly recovered within the present framework, with an even more simplified modelization in terms of just one main field, $n$, and one auxiliary one, $\phi$. Indeed, if we suppose electrons to be adiabatic, then $n \propto \exp(\phi)$. The white-noise ansatz on $\phi$ then yields after some algebra $P(n) \propto \exp\left[-(1/2)\left(\ln(n/n_0)/\sigma_n\right)^2\right] n^{-1}$.

A few words also about some of the limits of our treatment. Our postulates are doubtless insufficient when the correlations within the signals are important, say, when dealing with such quantities as statistics of laminar times or frequency spectra. By example, Eq. (5) is basically equivalent to a random walk and hence yields a $1/f^2$ frequency spectrum, that does not account for evidence from experiments. Our numerical simulations feature roughly a $1/f^2$ intermediate scaling and then a $1/f^4$ asymptotic beyond a crossover frequency $f^*$. Quite reasonably, the correlation time $t_{\text{corr}}$ of the fluctuations should be related to $f^*$: $t_{\text{corr}} \sim 1/f^*$, hence our simplified analytical model should be taken as an accurate picture of the dynamics over times scales larger than $1/f^*$.

The results of this work therefore unveil the reason for the universality of PDFs: non-Gaussian features arise because of non-linear interactions between the edge turbulent fields. However, precise details of the interactions are immaterial to a large extent, and may be replaced with phenomenological terms that are substantially the same for all kinds of instabilities, as long as they enter the equations in the form of quadratic couplings plus additive contributions. We considered here an instance of interchange instability, but it is clear that replacing it with other kinds of instabilities does not modify qualitatively the structure of Eq. (6). Furthermore, our results may be extended to other fields: we considered here density and vorticity only, but it is reasonable to expect that similar considerations hold, e.g., for temperature or magnetic field fluctuations: Indeed, we verified this statement by carrying on simulations with a three-fields model, in which also the equation for temperature is added to Eqns. (1,2), pretty much like done in [6,7], and thus the PDF for $T$ fluctuations may be computed.

This work was supported by the European Communities under the contract of Association between EURATOM/ENEA. S. Cappello helped to improve the manuscript. Comments from M. Ottaviani are acknowledged.